\documentclass[titlepage,letterpaper,twocolumn,prl,longbibliography,nolinenumbers]{revtex4-1}
\usepackage[utf8]{inputenc}
\usepackage{amsmath}
\usepackage{amssymb}
\usepackage{graphicx}
\usepackage{hyperref}
\usepackage[percent]{overpic}
\usepackage{pdfpages}
\usepackage[normalem]{ulem}
\usepackage[margin=1 in]{geometry}

\makeatletter
\AtBeginDocument{\let\LS@rot\@undefined}
\makeatother

\newcommand{\erf}{\textrm{erf}}

\begin{document}
\title{When Pull Turns to Shove: A Continuous-Time Model for Opinion Dynamics}

\author{David Sabin-Miller}
\email[]{\mbox{davidsabinmiller@u.northwestern.edu}}
\affiliation{Department of Engineering Sciences and Applied Mathematics, Northwestern University, Evanston, IL, USA}
\author{Daniel M.~Abrams}
\affiliation{Department of Engineering Sciences and Applied Mathematics, Northwestern University, Evanston, IL, USA}
\affiliation{Northwestern Institute for Complex Systems, Northwestern University, Evanston, IL, USA}
\affiliation{Department of Physics and Astronomy, Northwestern University, Evanston, IL, USA}

\begin{abstract}
Accurate modeling of opinion dynamics has the potential to help us understand polarization and what makes effective political discourse possible or impossible.  Here, we use physics-based methods to model the evolution of political opinions within a continuously distributed population.  We utilize a network-free system of determining political influence and a local-attraction, distal-repulsion dynamic for reaction to perceived content.  Our approach allows for the incorporation of intergroup bias such that messages from trusted in-group sources enjoy greater leeway than out-group ones.  We are able to extrapolate these nonlinear microscopic dynamics to macroscopic population distributions by using probabilistic functions representing biased environments.  The framework we put forward can reproduce real-world political distributions and experimentally observed dynamics, and is amenable to further refinement as more data becomes available.
\end{abstract}

\maketitle
\section{Introduction}
The field of opinion dynamics seeks to understand the evolution of ideas in populations, a complex interdisciplinary endeavor which has attracted a wide variety of approaches from different disciplines. After early mathematical groundwork \cite{holley1975infiniteVoter}, the growth of network science has led to a boom in models which utilize neighbor-based update rules to examine long-term outcomes for opinion distributions, such as polarization and consensus, e.g.~ 
\cite{deffuant2000mixing, hegselmann2002opinion,Sood2005voter, Castellano2009qvoter,hegselmann2005opinion, galam2007role, martins2008continuous, verma2014impact, Knopoff_2014, liu2014control,  wang2016bistability, DelVicario2017, pinasco2017modeling, afrasiabi2018opinion, Ye2020cont_multi, Alves2020prob_repel,kowalska2020grid_comp, Maciel2020multi_bound_conf}.
Other researchers have advanced ``sociophysics'' approaches, which apply techniques from statistical physics to analyze analogous social systems \cite{galam1982sociophysics, galam1986majority, galam1997rational, sznajd2000opinion, galam2004contrarian,  sznajd2005sznajd, toscani2006kinetic, bertotti2008discrete, Castellano2009statmech, during2009boltzmann, during2015opinion}.
Complementary to these modeling approaches, theoretical and empirical work from economics and social science has examined the political bias of media entities \cite{groseclose2005measure,entman2007framing} and their influence on a population \cite{Slater2007spiral, DellaVigna/Kaplan2006, Martin2017, acemoglu2011opinion, fan2017evolution}. All these approaches contribute valuable insight toward an understanding of this complex topic, but the disparities between their perspectives make direct cohesion a challenge.  

Our model takes a different approach, which we believe achieves the key benefits of previous models while expanding flexibility and retaining the ability to incorporate real-world data as it becomes available.  One key structural choice we make is to modularize the process of opinion change by breaking it into two parts: perceptions and reactions.

\begin{figure}[t!] 
	\centering
    \begin{overpic}[width=\columnwidth]{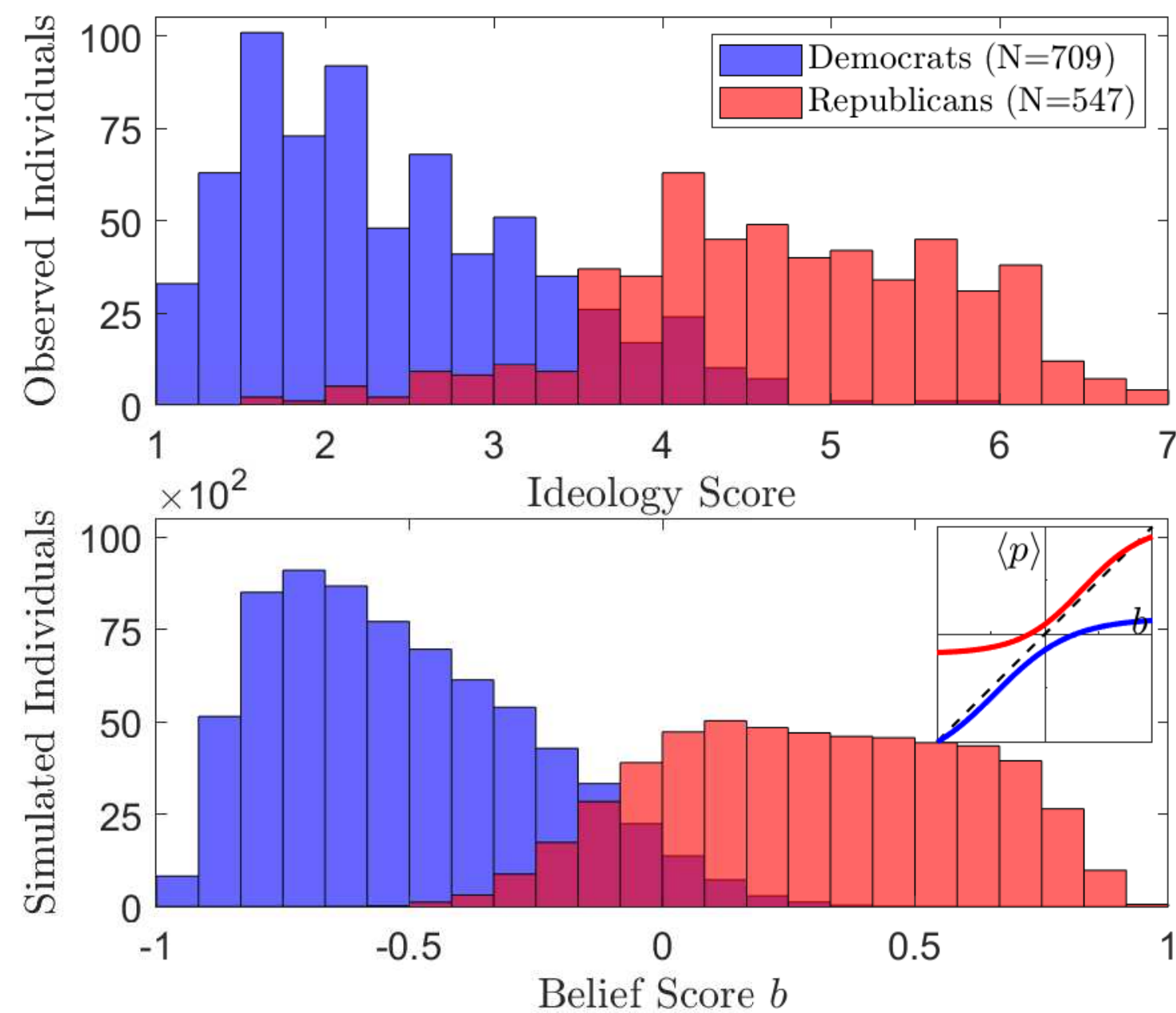}
        \put(15,80){\large{\textbf{a}}}
        \put(15,37){\large{\textbf{b}}}
    \end{overpic}
	\caption{\textbf{(a) Empirical ideological distributions by U.S.~political party.} Average ideological position score from 1 (strongly liberal) to 7 (strongly conservative) on social, economic, and military issues for 1256 U.S.~Twitter users. Data from \cite{Bail2018}. \textbf{(b) Model predictions.}  Steady state for our simulated population of 70,900 Democrats and 54,700 Republicans, with party perception curves shown in the inset. See Results section for details. }
	\label{fig:Bail_and_sim_hists}
\end{figure}

In our model, individuals perceive a probabilistic mix of politicized experiences which depends on their ideology and party. This might be thought of as the continuum limit of a network approach, where  influences are so numerous and varied that interactions are best characterized by a probability distribution rather than explicit neighboring agents.  This approach also allows us to encapsulate broader societal influences such as politicized media environments, since individuals' perceptual mix may be constantly changing to reflect their changing worldview.

We model individuals' reactions to these perceptions by having their ideology evolve in continuous time.  This is governed by ordinary and stochastic differential equations which depend on their current position and their perceptual distribution. 

Together, these perception and reaction modules capture a feedback loop between individuals' current beliefs, the biased ``slice'' of the political world they perceive, and how they update those beliefs as a result.

\subsection{Political Spectrum}
Like many prior approaches (e.g.~\cite{poole1984, deffuant2000mixing, hegselmann2002opinion, Knopoff_2014, DelVicario2017, DellaVigna/Kaplan2006, Martin2017,pew2017divide, groseclose2005measure, Bail2018}), we consider a single, finite ideology axis. This appears to be supported for the U.S.~political environment dominated by two parties: empirical results show that the liberal-conservative dimension captures the great majority of modern U.S.~legislative behavior \cite{hare2014polarization}. However, it's unclear how appropriate this assumption is for other countries with differing political systems: multiple dimensions may be warranted (see Possible Extensions section). 

Figure \ref{fig:Bail_and_sim_hists}(a) shows a one-dimensional projection of political ideology for the U.S.~population based on one study \cite{Bail2018}; though the precise methods of projecting the political landscape onto one axis differ between sources, other recent reports like that of Pew Research \cite{pew2017divide} show good qualitative agreement.

We will use the term \textit{belief score}, $b$, to refer to an individual's ideological position between $-1$ (extreme liberal) and $+1$ (extreme conservative).  We abstract all politically-opinionated information an individual is exposed to (hereafter termed \textit{percepts}, $p$) onto this same axis, so that a percept of $p=+0.5$ is in support of belief score $+0.5$ (conservative), a percept with value $p=0$ argues for a neutral stance, and so on. Due to the imprecise nature of any measurement on this scale (it's a projection of a highly abstract space that can be quantified in different ways), qualitative results should be robust to small changes in these values.

\subsection{Opinion Change}
Classic ``bounded-confidence'' models (e.g., \cite{deffuant2000mixing, hegselmann2002opinion}), which allow for individuals to interact only with others who are relatively like-minded, have been used to capture the effect of homophily on interaction.  But political issues are contentious and are often brought up between those who disagree, and are easily suffused with negative emotional affect rather than agreement or indifference.  Repulsion from disliked positions seems to be an important determinant in swing voters: a recent Pew survey \cite{pew2016animosity} found that U.S.~independents supporting one of the political parties did so mostly due to negative perceptions of the \textit{other} party. So like some other extensions to bounded-confidence models (e.g., \cite{DelVicario2017, Knopoff_2014}), we supplement local-attraction behavior with distal repulsion:  individuals who are exposed to ideas which are too different from their own will not be attracted, but rather be \textit{repelled} from the espoused position of the source.  There is experimental evidence that this can be a very potent and real source of ideological movement: in recent work from Bail et al.~\cite{Bail2018}, it was found that exposure to 24 tweets per day from prominent members of the opposing party can have a significant repulsive effect over the course of a month, even among all other political inputs received by the participants (self-identified politically active Twitter users).

\section{Methods}
The first key component of our model is the \textit{reaction function}. This is a continuous function which relates an individual's shift in ideological belief to the difference between a perceived political opinion (the percept, $p$) and the individual's own belief, $b$; we will refer to this difference $p-b$ as the \textit{dissonance}.  A repulsion effect will be modeled through the existence of a \textit{repulsion distance} $d$ such that percepts less dissonant than $d$ will be attractive and percepts more dissonant than $d$ will be repulsive.  This parameter $d$ can be allowed to vary depending on the context of the message, which will allow us to model the important effect of intergroup bias: for example, a somewhat challenging position can be repulsive when it comes from a disliked source but attractive when introduced by a member of one's in-group (see ``Adding Intergroup Bias'' below).

\begin{figure}[t!] 
	\centering 
	\includegraphics[width=\columnwidth]{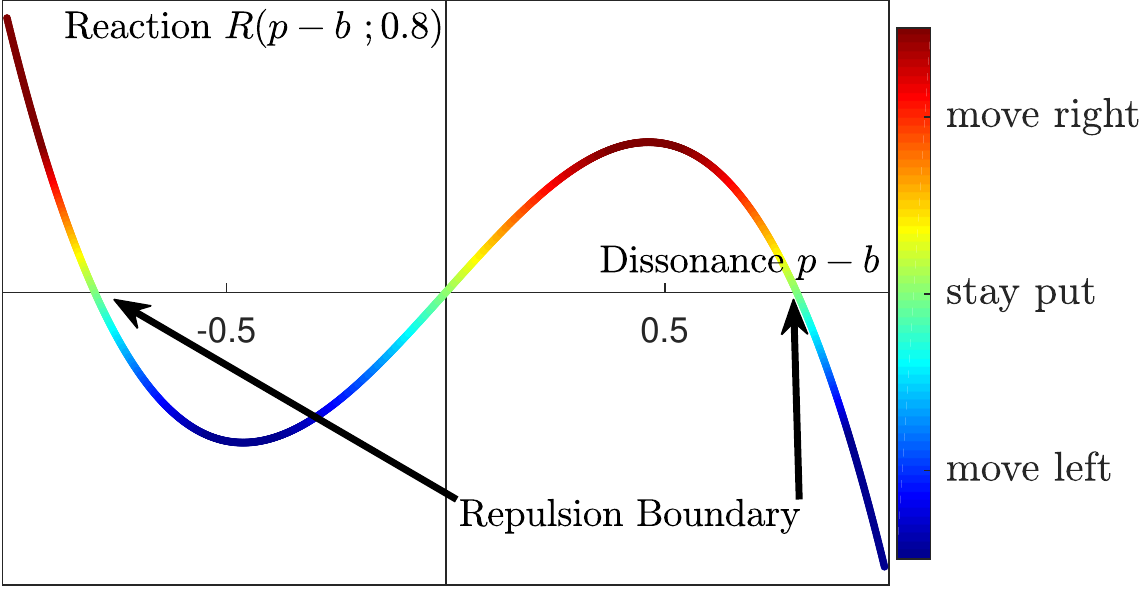}
	\caption{\textbf{Example reaction function.} Here we show a cubic reaction function, where an individual's reaction depends on dissonance $p-b$. Vertical scale has arbitrary units: the magnitude of this movement depends on time constant $\tau$ and current belief score $b$.  For this image a repulsion distance of $d=0.8$ was chosen. }
	\label{fig:reaction_func}
\end{figure}

One simple form for a reaction function that satisfies the above conditions employs a cubic dependence on dissonance: \begin{equation}
    R(p-b; d) = \left(p-b\right)\left[1 - \frac{(p-b)^2}{d^2}\right]\;,
    \label{eq:reaction}
\end{equation}
shown visually in Fig.~\ref{fig:reaction_func}. We utilize this cubic reaction function for all our models in this paper, but we expect similar results with any qualitatively similar function used in its place (ideally one inferred from experimental data). 


To organically constrain belief dynamics to a bounded domain (in our case, $[-1,1]$), we temper the above reaction function with a multiplicative factor $(1-b^2)$. This has the effect of gradually damping motion near the extremes---thus we interpret the $\pm 1$ boundaries of our finite ideology scale to be asymptotic extremes that are only approachable, not attainable. We also scale the dynamics by a time constant $\tau$ which controls the speed of belief change.  Then, for an individual $j$ with belief score $b_j$ and repulsion distance $d$, exposed to percept $p$ (which may depend on many factors), we arrive at the following differential equation for ideological dynamics:
\begin{equation}
    \tau \frac{\mathrm{d}{b_j}}{\mathrm{d}{t}} = (1-b_j^2) \left\{\left(p-b_j\right)\left[1 - \frac{(p-b_j)^2}{d^2}\right]\right\}\;.
    \label{eq:dbdt}
\end{equation}

\subsection{Perceptual Diets}
An important question remains: which individuals are exposed to which messages? The vast majority of work on opinion dynamics has been in a network context, wherein agents update their opinions according to a rule incorporating the positions of some other agent(s) (e.g.~\cite{galam2004contrarian, sznajd2000opinion,sznajd2005sznajd, holley1975infiniteVoter,Sood2005voter, Castellano2009qvoter, deffuant2000mixing, hegselmann2002opinion, hegselmann2005opinion, galam2007role, Castellano2009statmech, verma2014impact, Knopoff_2014, DelVicario2017, afrasiabi2018opinion}). Our approach sidesteps the need for constructing explicit influence networks, which are difficult to capture due to the many modalities of human interaction.  Instead we suppose that an individual's party affiliation and current political position determine their perceived ``slice'' of the political world---a probability distribution of political experiences, $\rho(p)$ \footnote{Note that this is not necessarily inconsistent with a network approach; we believe a dynamical interaction network dependent on affiliation and political position could lead to a similar model.}. This continuum approach allows us to personalize political environments to account for ``media bubbles'' and other biased environments even without a network, and is easily scaled to large populations.

\subsection{Toy Models}
\subsubsection{Simplest Model}
For the simplest concrete implementation of our framework, we might suppose a single-party population is initially distributed across the belief spectrum but is otherwise homogeneous, and that every individual perceives the same delta-distribution of political content, the constant percept $p = C$. Then upon choosing a repulsion distance $d$ we can exactly determine long-term behavior of the entire group---there will be a single flow function that affects the whole belief spectrum:
\begin{equation}
    \tau \frac{\mathrm{d}{b_j}}{\mathrm{d}{t}} = (1-b_j^2) \left\{\left(C-b_j\right)\left[1 - \frac{(C-b_j)^2}{d^2}\right]\right\}\;.
    \label{eq:constant_percept_flow}
\end{equation}
This ordinary differential equation (ODE) has fixed points at  $b_j = C,$ $b_j=C \pm d$, and $b_j = \pm 1$ (due to the imposed domain bounds). The fixed point at $b_j = C$ is stable, and stability of the other points alternates. 

\begin{figure}[t!] 
	\centering 
	\includegraphics[width=.84\columnwidth]{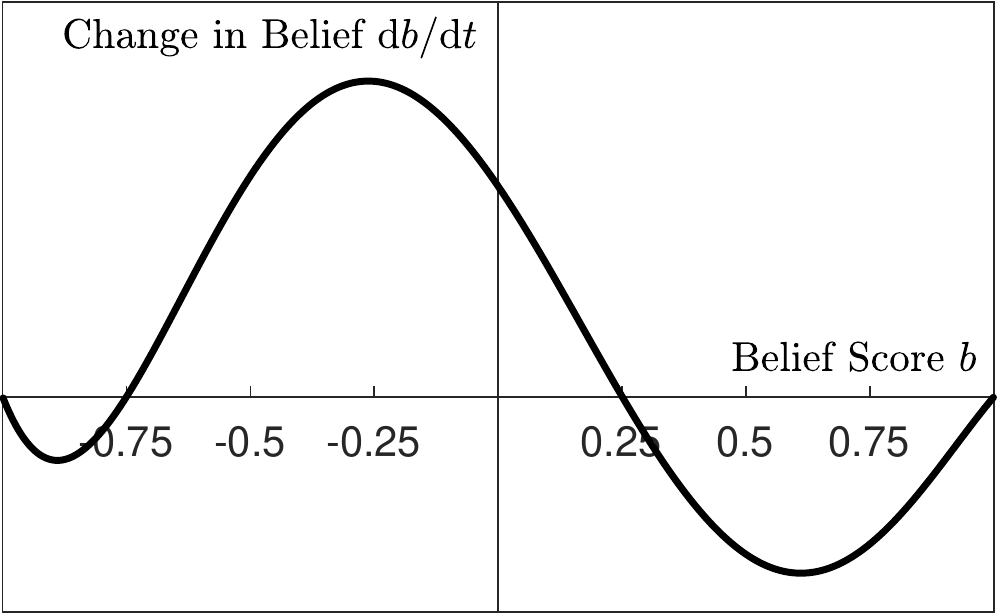}
	\caption{\textbf{Flow diagram.} Example of differential movement for a population uniformly exposed to a percept with score $+0.25$ assuming repulsion distance $1$ (see Eq.~\eqref{eq:constant_percept_flow}).  Vertical axis scaling is arbitrary. }
	\label{fig:rep_reaction}
\end{figure}

For example, if we use the cubic reaction function from Fig.~\ref{fig:reaction_func} above and set $d = 1$, $C = 0.25$, then that party's population experiences differential movement as shown in Fig.~\ref{fig:rep_reaction}.  Fixed points exist at $\{-1, -0.75, 0.25,  1, 1.25\}$ (though beliefs are constrained to the $[-1,1]$ domain, so the theoretical fixed point at $1.25$ is not meaningful).  Given time, all observers between $-0.75$ and $1$ would congregate at $0.25$, and all observers starting left of $-0.75$ would converge to $-1$. This small segment of the population---the members that are liberal enough to be repelled by the ``party line''---might be likely to switch parties in favor of one with more comfortable percepts, though we don't include such party-switching dynamics in this initial model.

\subsubsection{Adding Intergroup Bias}

We would also like our modeling framework to accommodate the tendency for individuals to be more receptive to information from those whom they perceive as allies, i.e., part of their ``in-group'' \cite{tajfel1979intergroup}. For the simplest case, we modify our previous model by adding an ``out-group'' with its own distinct constant ``party line'' percept $p_\textrm{o}$. Now percepts have a party identity attached to them, and we allow individuals to consume a mixed diet of in-group and out-group information, at belief scores of $p_\textrm{i}$ and $p_\textrm{o}$, respectively.  We set repulsion distances $d_\textrm{i}$ and $d_\textrm{o}$ for in-group (e.g., U.S.~Republican) and out-group (e.g., U.S~Democrat) messengers, with $d_\textrm{o}\leq d_\textrm{i}$. We can set a fixed fraction $f$ for in-party content, or allow for a belief-dependent skew $f(b)$ such that, e.g., liberal Republicans view a higher fraction of Democratic content than their conservative party-mates.   The average flow function $\textrm{d}b/\textrm{d}t$ is then a simple weighted average of the flow functions in Eq.~\eqref{eq:dbdt} due to each source:
\begin{equation}
    \tau \frac{\mathrm{d}{b}}{\mathrm{d}{t}} = (1-b^2) \left[ f R_\textrm{i} + (1-f) R_\textrm{o} \right]\;, 
    \label{eq:fractional content}
\end{equation}
where in general $f=f(b)$, $R_\textrm{i}=R(p_\textrm{i}-b; d_\textrm{i})$, and $R_\textrm{o}=R(p_\textrm{o}-b; d_\textrm{o})$.

\begin{figure}[t!] 
	\centering
	\includegraphics[width=\columnwidth]{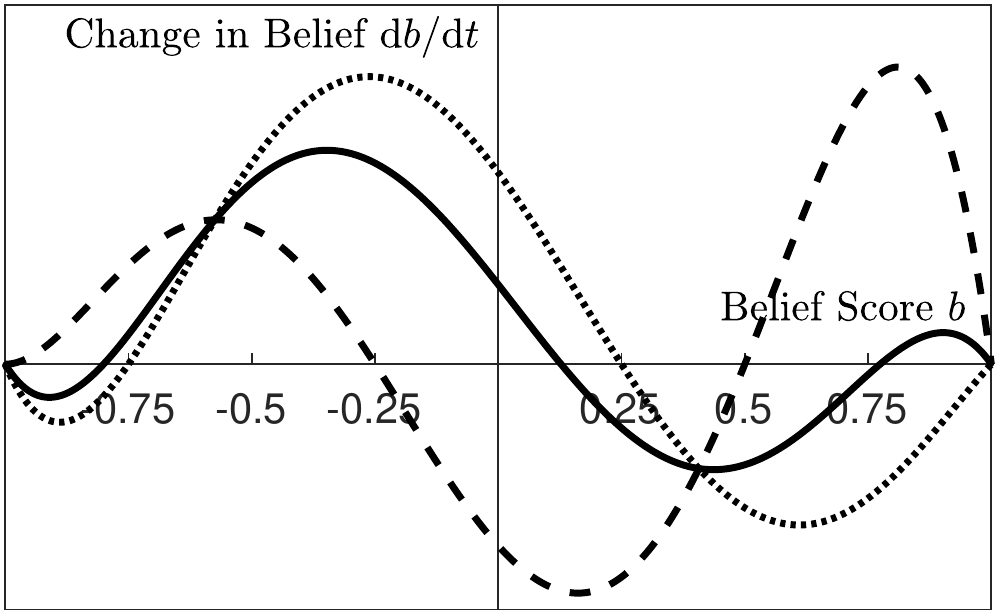}
	\caption{\textbf{Flow with different messengers.} The flow functions for in-group (dotted) and out-group (dashed) messages of $p_i=+0.25$ and $p_o=-0.25$ with repulsion distances of $1$ and $0.75$, respectively. The solid curve is the net flow if individuals are exposed to $70\%$ in-group and $30\%$ out-group percepts. Vertical axis scaling is arbitrary.}
	\label{fig:fractional}
\end{figure}

To understand the flow in this case, it is informative to consider the purely in-group and purely out-group situations ($f = 1$ or $0$, respectively), because all fractional perceptual ``diets'' are interpolated between them (see Fig.~\ref{fig:fractional}).  We note that exposure to some out-group content can in some cases \textit{increase} polarization for a small extreme group---for example, in Fig.~\ref{fig:fractional}, individuals starting with $b>0.75$ will on average move \textit{rightward} when exposed to percepts from a $70\%/30\%$ combination of in-group and out-group sources, respectively (solid curve), whereas those same individuals would move \textit{leftward} if presented with in-group information alone (dotted curve).   This simple example shows how exposure to---and rejection of---opposing content can have a polarizing influence on a population.

Note that we assume that this ``tribal'' bias only affects the \textit{reaction} to content, not its subjectively perceived ideological score $p$. However, the inclusion of such an additional bias effect is reasonable, and may be handled with a slight increase to model complexity (see Possible Extensions section). 

\subsubsection{Adding Personalized Perceptions}
Putting aside the in-group/out-group balance for a moment, we might expand our simplest model in a different way: by linking individuals' perceptions to their current beliefs via a ``perception curve'' $p(b)$, which indicates what content they see as a function of position. This reflects the differing ``slices'' of the political world that individuals see as a result of the differing environments and personal biases that accompany their ideologies.

In our simplest model, where $p=C$, the perception curve is a horizontal line in $b$ vs $p$ space; individuals at all $b$ values perceive the same thing.  In a hypothetical ``perfectly targeted'' world, the perception curve would be the $45^{\circ}$ line $p=b$, and nobody would change belief because each person would perceive content perfectly in line with their current worldview.

Luckily, we don't need to privilege one such curve in particular---a graphical analysis method lets us combine \textit{any} perception curve with the reaction function and read off a (qualitative) flow for each segment of the population.  To do this, we plot the perception curve $p(b)$, and overlay the $45^{\circ}$ line for reference---any time the perception curve intersects it, the individuals at that belief score are stationary, since their perceptions are in agreement with their current beliefs.  If the perception curve is slightly above the $45^{\circ}$ line, individuals with those beliefs are perceiving something slightly more conservative than their own views, and move right.  Similarly, people move left wherever the perception curve is slightly below the $45^{\circ}$ line.  

We also overlay the repulsion boundaries at distance $d$ above and below that $p=b$ line.  If the perception curve exits the resulting ``trust band'' over some $b$ interval, that segment of the population is repelled and moves the opposite direction from what would be expected based on small deviations from the $45^{\circ}$ line.  

\begin{figure}[t!] 
	\centering 
	\includegraphics[width=\columnwidth]{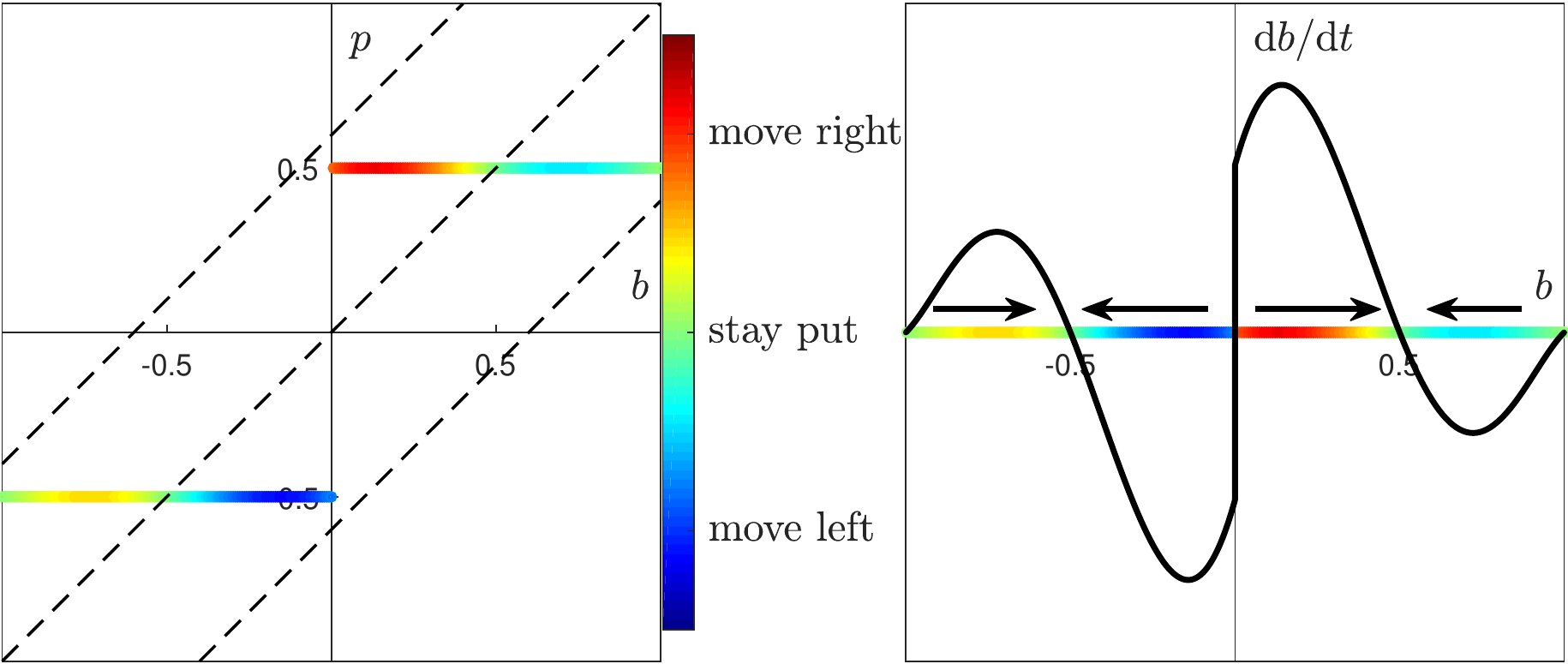}\\
	\includegraphics[width=\columnwidth]{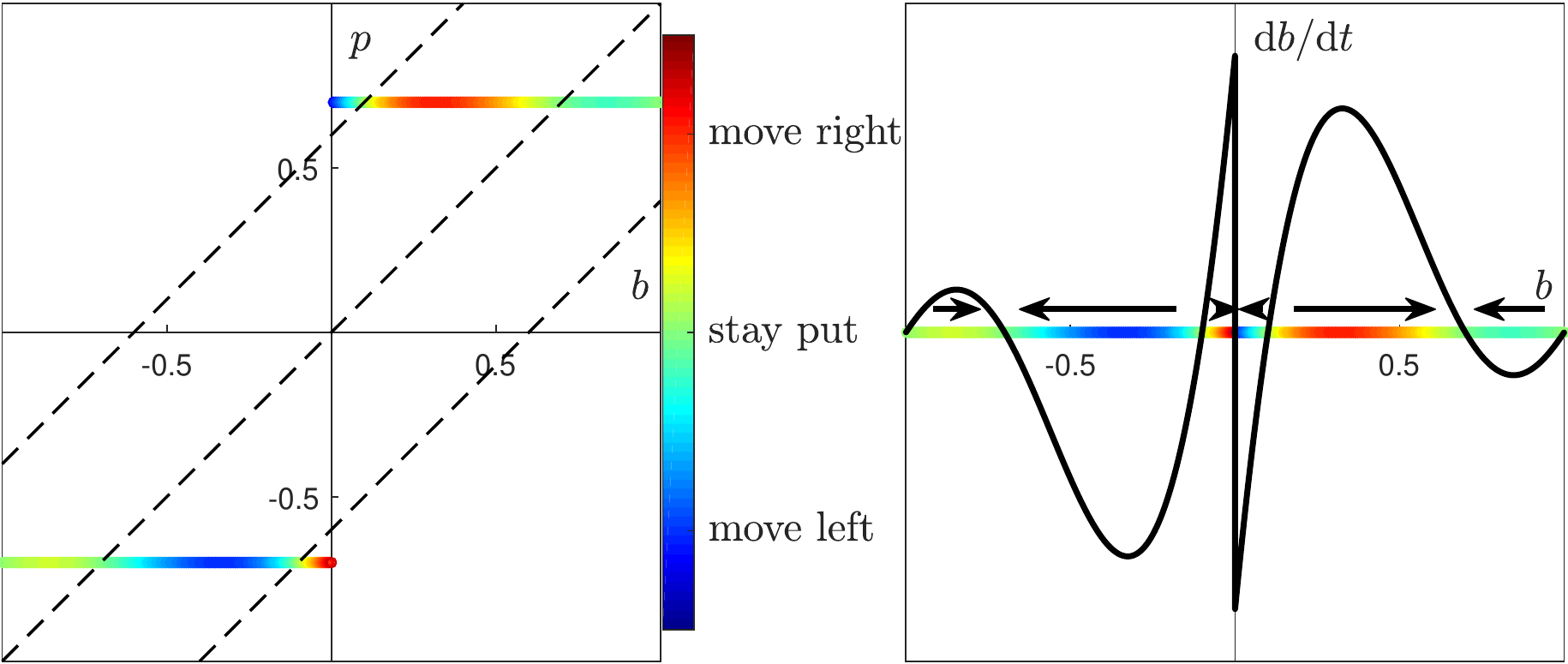}
	\caption{\textbf{Two-take world.} Graphical analysis of step-function perception curves.  Left panels: perception curves color coded for the movement induced,  along with dashed $p=b$ line and repulsion boundaries.  Right panels: Projection of that flow-velocity color onto the belief axis, compared with the exact population flow calculated from Eq.~\eqref{eq:dbdt} (black curve).  Top row: When perception curves lie within the trust region, we see two attractors at the ``party line'' belief values. Bottom row: with more extreme ``party lines,'' centrists are repelled by either party position, creating a stable central attractor. }
	\label{fig:two_take_doubles}
\end{figure}

It is then straightforward to determine the qualitative behavior of the whole population given any perception curve $p(b)$ by visually examining intersections of the perception curve with the $45^{\circ}$ lines, as in Fig.~\ref{fig:two_take_doubles} (left panels). With a closed form expression for $p(b)$, we can use Eq.~\eqref{eq:dbdt} to obtain an exact flow function (black curves on right panels of Fig. \ref{fig:two_take_doubles}), and confirm our qualitative analysis.  If multiple parties are present, this analysis is performed separately for each, and the resulting reactions are combined as in Eq.~\eqref{eq:fractional content}.

The real benefit of this graphical approach is its generality; one can draw any perception curve one would like and simply read off the fixed points and stability.  Whenever the perception curve crosses the diagonal with slope less than one, that crossing becomes a stable fixed point.  Whenever it crosses with a slope greater than one, the crossing becomes an unstable fixed point instead.  If the perception curve crosses a repulsion boundary, shallow crossings create unstable points and steep crossings create stable ones.  

While the choice of perception curve entails a large degree of modeling freedom,  based on our graphical analysis reasoning we know our model's qualitative predictions aren't particularly sensitive to the choice. Ideally, real-world data could (and should) be used to construct such a curve (e.g., by evaluating the partisan positions of news sources and other political influences experienced by individuals across the political spectrum), though we leave this for future work (see Future Work subsection below).

\subsubsection{Adding Heterogeneity}
\begin{figure}[t!]
	\centering 
	\includegraphics[width=\columnwidth]{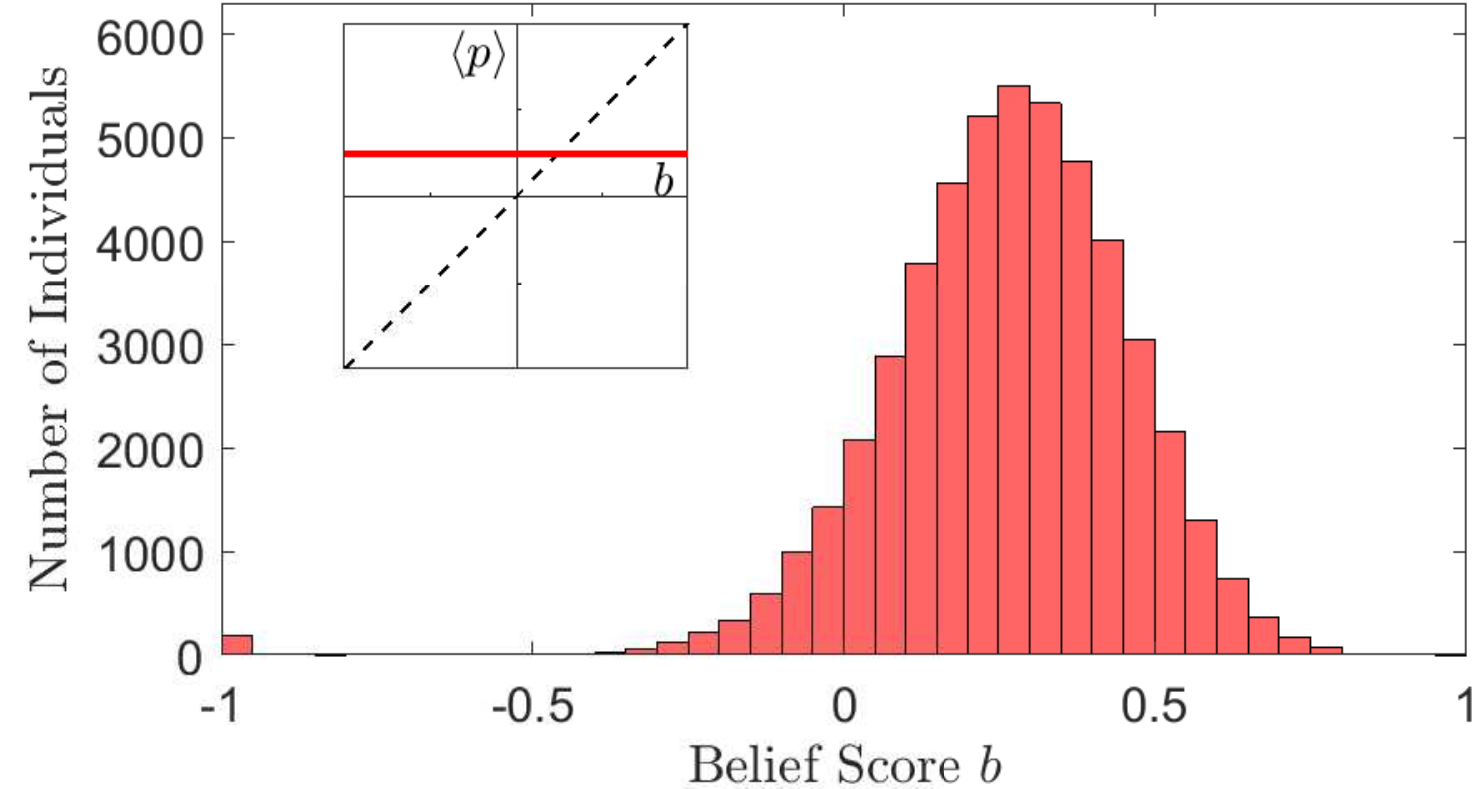}
	\caption{\textbf{Simulated population distribution.} The stable population state induced by flow function Eq.~\eqref{eq:constant_percept_flow} with added Gaussian noise ($\tau = 1$, $dt = 0.001$, $\sigma = 0.25$, $N=50,000$). Population was initialized to match Republicans from Bail \cite{Bail2018}. Inset: Perception curve, the constant $C=0.25$. All variation is in reaction. }
	\label{fig:rep_dist}
\end{figure}

To move toward a more realistic scenario, we must allow for heterogeneity of both environments and individuals.  We can introduce random variation in two distinct components of the model: perceptions (so individuals are exposed to a range of different inputs rather than a single determined value), and the reaction function (so otherwise identical individuals can react differently to the same percept).  For the latter, we add Gaussian noise to the reaction function $R$, which causes the stable fixed points from our prior analysis to expand into finite-width stable distributions; these may be estimated easily and accurately by Euler-Maruyama numerical integration of our now-stochastic differential equation (SDE).  For example, with the conditions in Fig.~\ref{fig:rep_dist}, the main body of the party congregates around the primary attractor at $0.25$, and a small group is repelled to $-1$ \footnote{In cases like this with multiple attracting ``camps'' without significant overlap, the long-term populations of each camp may depend on initial conditions. }.

If we wish to add variability to the percept instead of the reaction, nonlinear effects become more important, since $p$---now properly a probability distribution $\rho(p)$---must be fed through reaction function $R(p-b)$ before its effects are determined. We suppose that percepts occur on a significantly faster time scale than opinion change itself, so that the net effect of the perceptual diet is a weighted average over all possible percepts, which for smooth percept distributions becomes an integral of $R(p-b)$ against $\rho(p)$.

Regardless of the shape of perception distribution, we can still use our graphical analysis technique to solve for net opinion drift, though the repulsion boundaries may warp, as we show in Appendix A.

\subsubsection{Full Model}
Taking all these effects together, the model has the following structure: 
\begin{subequations}
\begin{gather}
R(p-b;d) = (p-b)\left[1-\frac{(p-b)^2}{d^2}\right] \label{eq:reaction2}\\
v_{in} = \int_{-1}^{1}R(p-b;d_{in})\ \rho_{in}(p ;b,\sigma_p) \textrm{d}p \label{eq:v_in}\\
v_{out} = \int_{-1}^{1}R(p-b;d_{out})\ \rho_{out}(p ;b,\sigma_p) \textrm{d}p \label{eq:v_out}\\
\tau\ \textrm{d}b = (1-b^2) \left\{ \left[f v_{in} + (1-f)v_{out} \right] \textrm{d}t + \sigma_r \textrm{d}W \right\} \label{eq:SDE}
\end{gather}
\end{subequations}  
with $\rho_{in}$ and $\rho_{out}$ as the perceptual distributions for in-group and out-group content respectively (with standard deviation $\sigma_p$), and $v_{in}$ and $v_{out}$ the opinion drift due to those influences.

This framework allows examination of the long-term impact of a probabilistic content-generating environment---which differentiates based on party and ideology---on a population which reacts to what they experience in a nonlinear manner heavily influenced by identity.

\section{Results}
The full model, with some reasonable parameter assumptions, exhibited equilibrium distributions which agree well with real-world observations (see Fig.~\ref{fig:Bail_and_sim_hists}).  
To represent a ``media bubble'' effect, we scaled in-group fraction linearly and symmetrically from $0.5$ (for $b=+1$ Democrats and $b=-1$ Republicans) to $0.9$ (for $b=-1$ Democrats and $b=+1$ Republicans):
\begin{subequations}
\begin{align}
    f_D(b) &= 0.7+0.2b, \\
    f_R(b) &= 0.7-0.2b\;.
\end{align}
\end{subequations}
Equations \eqref{eq:reaction2}, \eqref{eq:v_in}, \eqref{eq:v_out}, \eqref{eq:SDE} determined population movement over time, simulated using Euler-Maruyama numerical integration.
Perceptual diets were beta distributions bounded by [-1,1], with standard deviation $\sigma_p = 0.2$ and peak (mode) given by sigmoid perception curves $p_D$ (for Democrats) and $p_R$ (for Republicans):
\begin{subequations}
\begin{align}
p_D(b) &= 0.7 \tanh\left[\frac{1.00}{0.7} (b+0.46)\right] - 0.55,\\
p_R(b) &= 0.6 \tanh\left[\frac{1.05}{0.6} (b-0.35)\right] + 0.42\;.
\end{align}
\end{subequations}
For details on these beta distributions and the implementation of equations \eqref{eq:v_in} and \eqref{eq:v_out}, see Appendix A3 and Appendix B.
Reaction parameter values were
$d_i = 1.3$, $d_o = 0.8$,  $\sigma_r = 0.15$.
%
For finding equilibria, the time constant $\tau = 1$ was used, and populations of both parties were initialized as uniformly distributed on $[-1,1]$. 
 
For easy comparison with real data, Fig.~\ref{fig:Bail_and_sim_hists}(b) shows a simulation of one hundred times Bail et al.'s experimental population: 70,900 Democrats and 54,700 Republicans. In this comparison, we must note that our belief scale is not identical to theirs; $\pm 1$ on our scale are asymptotically extreme, whereas 1 and 7 on Bail et al.'s scale are attainable and signify strong agreement on all surveyed issues.

We also replicated the experiment of Bail et al.~in silico: starting with a population at equilibrium (shown in Fig.~\ref{fig:Bail_and_sim_hists}(b)), and artificially inducing counter-attitudinal Democratic content to Republican experimental subjects (a beta distribution peaked at $p=-0.75$, weighted as if it consisted of 24 percepts on top of a presumed diet of 100 percepts per day) over the course of 30 ``days,'' caused the mean position of those subjects to shift rightwards by a little less than half its natural standard deviation (from $0.30$ to $0.42$, stdev $\sigma \approx 0.3$). This matches the findings of Bail et al., who found average rightward movement of $0.6$ points on a 1-to-7 scale, which represented between $0.11$ and $0.59$ standard deviations ($p<0.01$) \cite{Bail2018}. 
Further implementation details can be found in Appendix B.

\section{Possible Extensions}
In order to keep the number of tune-able parameters and functional forms to a minimum in the absence of much constraining data, we have made many simplifying assumptions. As such, it is easy to imagine extensions which might increase the realism of this model, and which would ideally be implemented when data is available to constrain them. 

One simple extension is the addition of more groups/parties, such as independent/unaligned individuals and messages. This would require another perception curve per group, and a three- or more-way fractional content breakdown instead of the single in-group fraction $f(b)$ as our analysis used. Different levels of out-group trust would be represented by different repulsion distances for each type of source.

Additional affiliations beyond party, such as religion, race, regional identity, etc., could also be added to the model. However, the number of identity combinations rises exponentially 
with each additional affiliation type, and each could affect the perception curve(s)---since affiliations can change the environment individuals are exposed to---and inter-group trust levels, so we submit that this option should be approached with caution. 

The repulsion distance $d$, representing ``trust,'' ``credulity,'' or ``benefit of the doubt,'' need not be binary or universal across the population.  Especially in the presence of multiple identity markers, one might let $d$ for each interaction depend on each group identity of the individual and of the messenger to allow for more intricate inter-group prejudices. One could also add noise to $d$ values to model individual variation in level of credulity towards other groups, or let $d$ evolve dynamically over time or as a function of political position. However, given the difficulty of measuring inter-group trust levels, we chose to avoid over-fitting by using only two $d$ values (in-group and out-group) in our simulations.

One might also reasonably suggest that the political content representing each party could depend on an observer's own party, not just their ideological position $b$. For instance, a liberal Democrat might perceive different Republican content than a liberal Republican. This would require additional perception curves, rather than re-using the same curve for all observers---each observer/messenger identity pair would require its own curve. This would allow for the possibility that individuals perceive a more extreme version of the other party as they become more extreme themselves, in other words a \textit{negatively} sloped perception curve for out-group content.  We note that the large dissonance numbers likely in this case would require a reaction function in which repulsion saturates (i.e., not our cubic assumption) to avoid dwarfing the effect of attraction.

As some other models have attempted \cite{Ye2020cont_multi,Maciel2020multi_bound_conf}, one might also consider multiple ideological dimensions: instead of a scalar belief value $b$, an $n$-dimensional vector $\mathbf{b}$ would represent an individual's beliefs with respect to each of $n$ issue axes. Percepts would engage with one or more of these issues.  Lacking relevant data, we do not put forward assumptions on how reaction dynamics might be coupled; one might assume that dynamics along each axis would be largely independent of one another, since position on one issue rarely affects position on another directly. However, it is possible that the dynamics along multiple axes would be coupled by tribalism; being repelled from a message might drive an individual closer to the opposing camp on more axes than just the one being engaged with, as the individual identifies more strongly with the whole opposing party.

Finally, one might add mechanisms by which the perception curves can change over time. Time-dependence could be introduced to investigate hypotheses about the impact of changing or time-correlated media environments, or perception curves might evolve in response to the population state.  The latter option would provide a form of indirect coupling between modeled individuals, allowing it to be self-contained over long time scales, but would require conjecture about how environment-creating entities perceive the population state and strategize in response.
  
\section{Discussion}

We have put forward a modeling framework for individual political opinion drift which separates perceived content and the reaction of the viewer to that content, in order to separately model perceptual filtering, the shift from attraction to repulsion for dissonant content, and the effect of intergroup bias.  We have presented toy models to elucidate each effect on its own in the absence of noise, and introduced a graphical analysis technique for qualitative analysis of behavior under general belief-dependent perception curves. With the inclusion of additive noise, analytically determined fixed points widen into stable distributions.
  
With all these effects included and some simple parameter assumptions, we showed that population distributions matching recent survey data emerge naturally.  Furthermore, we were able to simulate the experiment of Bail et al.~\cite{Bail2018} and found similar dynamics under counter-attitudinal perturbation.  

\subsection{Further Work}
A paucity of available data has forced us to make assumptions on functional forms and parameter values. While these are reasonable placeholders, they can be modified or replaced as empirical data become available; it isn't hard to imagine experiments which might elucidate qualitative and quantitative effects of interest.

For example, to refine the reaction function, further experiments like that of Bail et al.~\cite{Bail2018} might investigate the impact of political opinions on individuals, and how the messenger's apparent identity affects the reception of dissonant ideas.

The lack of reliance on network structure means that data collection can focus on averages and distributions rather than influence-network properties and tie reconstruction.  Perceptual diets might be estimated from the top down, by assigning each media outlet or other notable source of political influence an ideology score (as others have done, e.g., \cite{Martin2017, groseclose2005measure, entman2007framing, chart}), and surveys or viewership data could determine which content is consumed in what proportion by each part of the ideological spectrum. Alternatively, self-report of political influences and their positions could produce estimates of perceptual diets which also account for interpretation bias---the same content might be interpreted differently by different observers.

Overall, we believe this model shows considerable promise in replicating dynamics and distributions from the real world. We have intentionally chosen a relatively simple structure which is nonetheless able to capture important psychological tendencies for repulsion and tribalism, and couple them to a politicized environment, while preserving mathematical tractability. However, as outlined above, the realism can be significantly increased with extensions which augment the framework.  

As data are collected to inform the base model and extensions, the validity and predictive power will grow. In this way, we hope that this framework will offer a lens with which to better understand individual and population-level opinion dynamics, and the feedback effects that arise due to the modern reality of personalized political environments.  We hope this endeavor leads to a new sort of data-driven political modeling to better understand human behavior, polarization, and strategies for effective political dialogue.


\section{Acknowledgments}
\begin{acknowledgments}
The authors would like to acknowledge NSF support through the Graduate Research Fellowship Program. The authors also thank Profs.~Niall Mangan and Hermann Riecke for useful comments and feedback.
\end{acknowledgments}

DSM proposed the initial model and performed numerical simulations.  DMA contributed to modifications of the model.  Both authors participated in writing of the manuscript.

The authors have no competing interests.

All data and code is available (see Appendix B).


%

\appendix
\setcounter{equation}{0}
\setcounter{figure}{0}
\renewcommand{\theequation}{A\arabic{equation}}
\renewcommand\thefigure{A\arabic{figure}} 
\onecolumngrid
\section{Appendix A: Perception Distributions}
As we mention in the ``adding heterogeneity'' section of the main paper, if our model is to have any claim at accurately modeling the political lives of real people, it must allow individuals to consume not just a single, constant percept $p(b)$ but rather a whole distribution of content, $\rho(p ; b,\sigma_p)$. In this case, instead of using the single $p$ value to determine an individual's reaction, we calculate their weighted-average reaction by integrating the probability distribution of percepts they might receive multiplied by the reaction those percepts would cause.  We note that our cubic reaction function is asymmetric across the repulsion boundary (it's steeper outside the boundary than inside, so repulsion is ``stronger'' than attraction). Thus, if individuals receive a distribution of percepts centered at their ``perception curve'' value $p(b)$, a symmetric widening of their experiences has the \textit{asymmetric} effect of shifting the system's fixed points: since it takes fewer repulsive events than attractive ones to maintain net-zero movement, the new fixed point occurs when the center of the perceptual distribution is still in the trust region. In other words, the repulsion boundary is effectively narrowed with regard to the peak percept value $p(b)$.  

In Appendix A we will focus on opinion drift caused by perceptions from a single group, neglecting reaction noise (i.e.~$\tau\ \textrm{d}b/\textrm{d}t=(1-b^2)v$, a simplification of equation \eqref{eq:SDE}), but dynamics in a full model are a simple linear combination of two such sources and a reaction-noise term.

The precise effects of perceptual variety depend on the shape of the perceptual distribution and the choice of reaction function.  We now consider three options in order of increasing realism.

\subsection{1. Gaussian Distributed Percepts}
First we'll consider Gaussian-distributed percepts centered on the ``perception curve'' value $p(b)$, and the cubic reaction function from Eq.~(1) of the main text, $(p-b)[1-(p-b)^2/d^2]$.  These choices are convenient in that the integral for average belief change is analytically tractable.  For clarity, we change variables to ``average dissonance'' $\mu = p(b)-b$, and let $x$ be the dummy variable of integration for possible dissonance.  If we allow percepts outside of $[-1,1]$ in this way, the integral is quite clean: 
\begin{align}
    \tau  \frac{\textrm{d}b}{\textrm{d}t}  &= 
    \left(1-b^2 \right) \int\displaylimits_{-\infty}^{+\infty}  
    \underbrace{ x\left(1 - \frac{x^2}{d^2}\right)}_{R(x;d)} 
    \underbrace{\left[ \frac{1}{\sqrt{2\pi} \sigma_p} e^{-\frac{(x-\mu)^2}{2\sigma_p^2} }  \right]}_{\rho(x;b, \sigma_p)}
    \textrm{d}x \nonumber \\
     &= \left(1-b^2 \right) \ \mu\left[ \frac{(d^2-3\sigma_p^2)}{d^2} - \frac{\mu^2}{d^2} \right] \;.
    \label{eq:approx_integral}
\end{align}

\begin{figure}[thp!] 
	\centering 
	\includegraphics[width=0.4\columnwidth]{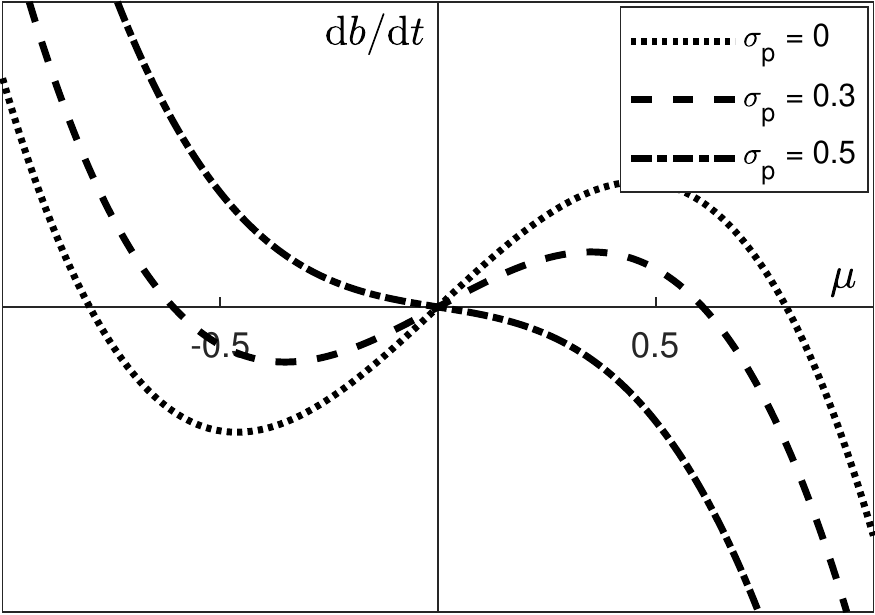}
	\caption{\textbf{Effect of perception distribution width on reaction.}  Net change in belief $\textrm{d}b/\textrm{d}t$ versus expected value of dissonance $\mu$ for varying levels of perception distribution width, from Eq.~\eqref{eq:approx_integral} with $d = 0.8$.    The critical standard deviation for $d=0.8$ is $\sigma_c = 0.8/\sqrt{3} \approx 0.46$. }
	\label{fig:bifurc_react}
\end{figure}

This is a cubic in $\mu$ with zeros representing the three fixed points: at $\mu=0$, and at $\mu = \pm\sqrt{d^2 - 3\sigma_p^2}$. So as the parameter $\sigma_p$ grows, we see a pitchfork bifurcation as the non-origin zeros disappear: for $\sigma_p > \sigma_c = d/\sqrt{3}$, the bracketed term in Eq.~\eqref{eq:approx_integral} is always negative. This means the net movement of the individual is \textit{away} from the average percept they see. See Figure \ref{fig:bifurc_react}.

These distribution widths are not unrealistically large; as seen in Fig.~\ref{fig:bifurc_react}, for a repulsion distance of $0.8$ the standard deviation needs only be $0.46$ for the overall effect of a content distribution to be repulsive (i.e.~causing movement away from that distribution's mean).  Thus, especially for out-group content with a naturally narrower repulsion distance, viewing a wider distribution of that content can actually cause repulsion, since the extreme percepts will repel the viewer more than the moderate percepts will attract them.

\begin{figure}[ht!] 
	\centering 
	\includegraphics[width=0.63\columnwidth]{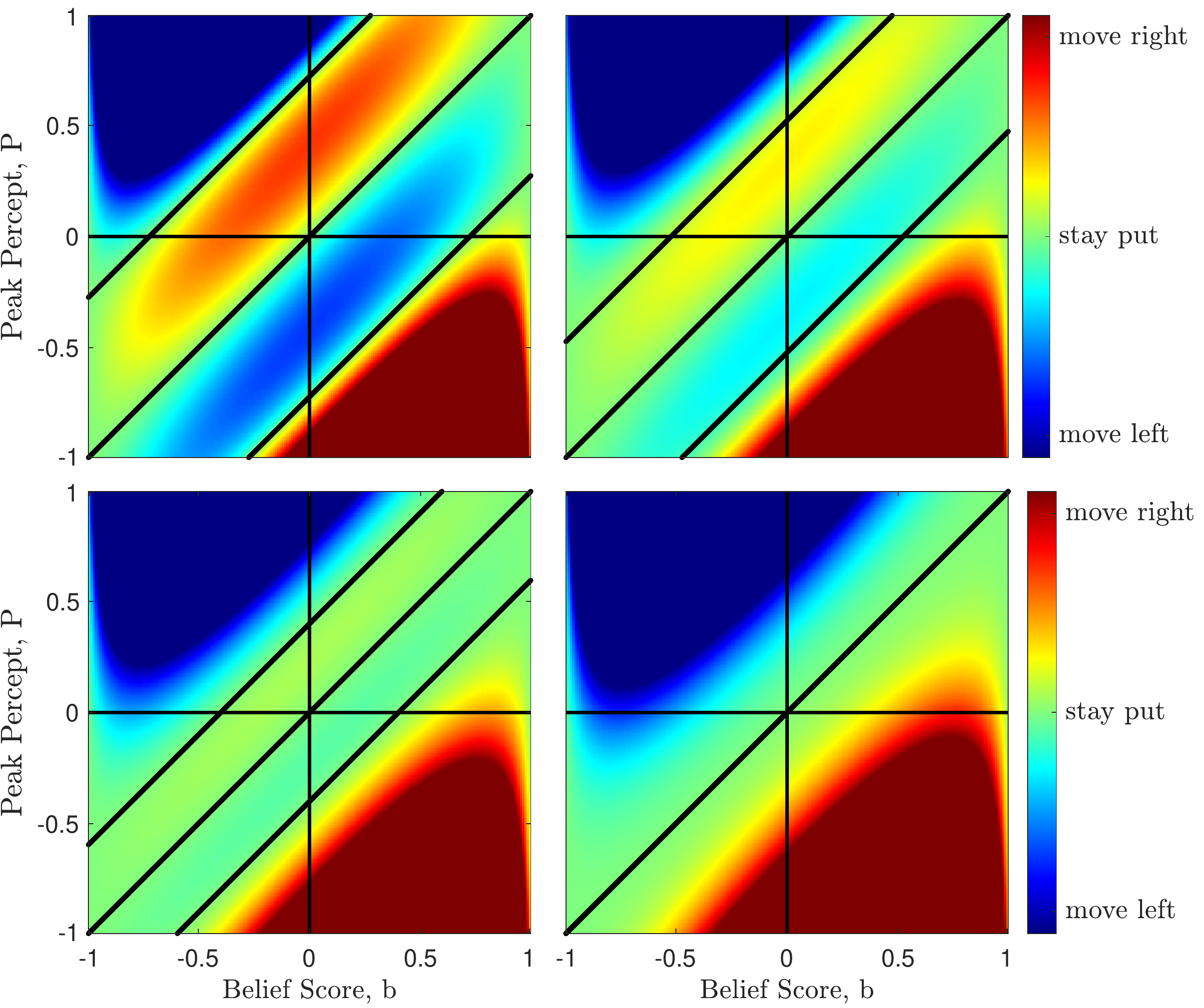}
	\caption{\textbf{Reaction map for normally distributed diets.} Net movement caused by normally distributed perceptual diets with peak $P = p(b)$, for individuals at belief score $b$, and repulsion distance $d = 0.8$. Results shown for $\sigma_p = 0.2$ (top left),  $0.3$ (top right), $0.4$ (bottom left), and $0.5$ (bottom right).  }
	\label{fig:unbdd_react_map}
\end{figure}

To visualize the effects of normally distributed perceptual distributions $\rho(p;b, \sigma_p)$ replacing deterministic percepts $p(b)$, we can examine density plots for the net movement for all combinations of $b$ and $p$ (repulsion distance $d=0.8$): see Fig.~\ref{fig:unbdd_react_map}. This is the space that our graphical analysis technique utilizes: if we establish a perception curve $p(b)$, the values of this map that the curve crosses are the realized average movement for each part of the population.

\pagebreak

\subsection{2. Truncated Gaussian Percepts}

In deriving Eq.~\ref{eq:approx_integral} we approximated by integrating over the entire real line for dissonance when it should be constrained to the range allowed by percepts in $[-1,1]$---that is, from dissonance $x = -1-b$ to $1-b$. That makes the result somewhat more complicated (note: lacking symmetry around $b$, we don't utilize the $\mu$ substitution, and $x$ represents  percept value instead of dissonance):
\begin{equation*}
    \tau \frac{\mathrm{d}b}{\mathrm{d}t}  = (1-b^2) \int_{-1}^1  \left(x-b\right)\left[1 - \frac{(x-b)^2}{d^2}\right] \left[ \frac{A}{\sqrt{2\pi}\sigma_p} e^{-\frac{(x-P)^2}{2\sigma_p^2} }  \right]  \textrm{d}x 
\end{equation*}
\begin{align}
    = A\frac{\left(1-b^2 \right)\sigma_p}{d^2 \sqrt{2 \pi}} &\Bigg\{ \left(1 - d^2 + 2 \sigma_p^2 + P^2 -3 P b + 2 b^2  \right) \left[ e^{\frac{-(-1+P)^2}{2 \sigma_p^2}} - e^{\frac{-(1+P)^2}{ 2 \sigma_p^2}} \right]  \nonumber\\
    & \qquad \qquad \qquad \qquad \qquad \qquad \qquad \qquad \qquad +\left( P-3 b\right) \left[ e^{\frac{-(-1+P)^2}{ 2 \sigma_p^2}} + e^{\frac{-(1+P)^2}{ 2 \sigma_p^2}} \right] \Bigg\}  \nonumber \\
    & + \frac{\left(1-b^2\right) (P-b) }{2d^2} \left[d^2-3\sigma_p^2-\left(P-b\right)^2 \right]\left[ \erf \left(\frac{1+P}{\sqrt{2} \sigma_p} \right) + \erf \left(\frac{-1+P}{\sqrt{2} \sigma_p} \right) \right]\;,
    \label{eq:true_integralpb}    
\end{align}
using shorthand $P = p(b)$ for compactness. 
We also note that $\sigma_p$ in this case is the standard deviation of the full Gaussian, not the truncated one. 
$A$ is a normalization factor depending on $b$ and $\sigma_p$ needed to make the truncated Gaussian integrate to 1:
\[
A = \frac{1 }{\frac{1}{\sqrt{2 \pi} \sigma_p} \int_{-1}^{1} e^{\frac{-(x-P)^2}{2\sigma_p^2}} \textrm{d}x } = \frac{2}{\erf \left( \frac{1+P}{\sqrt{2}\sigma_p} \right) - \erf \left( \frac{-1+P}{\sqrt{2}\sigma_p} \right)}
\]

\begin{figure}[h!]
	\centering 
	\includegraphics[width=0.63\columnwidth]{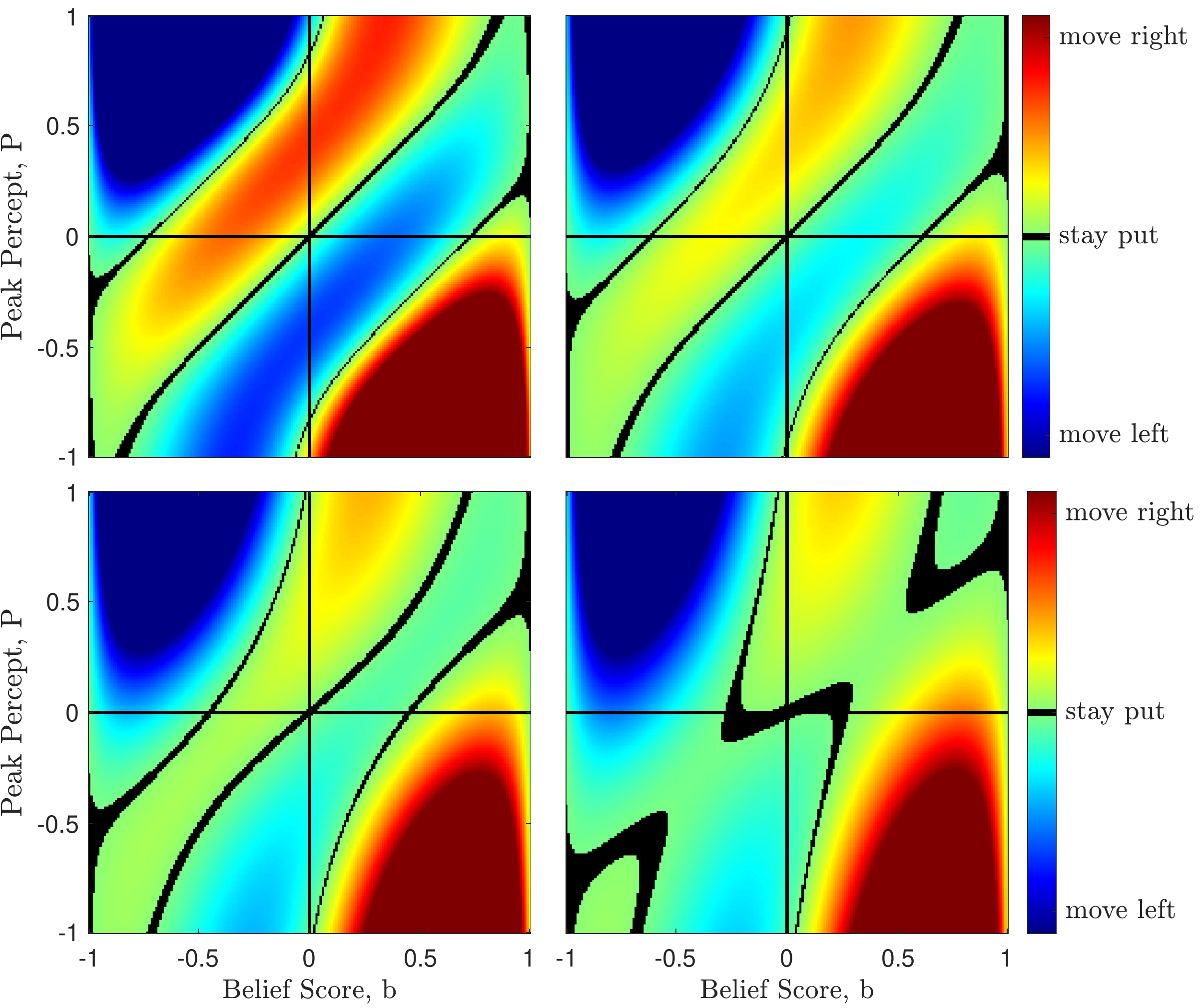}
	\caption{\textbf{Reaction map for truncated-normal diets.} Net movement caused by truncated-normal perceptual diets with peak $P$ (vertical axis), for individuals at belief score $b$ (horizontal axis), with repulsion distance $d = 0.8$. Results shown for $\sigma_p = 0.2$ (top left),  $0.3$ (top right), $0.4$ (bottom left), and $0.5$ (bottom right).}
	\label{fig:trunc_norm_react_map}
\end{figure}

 Fig.~\ref{fig:trunc_norm_react_map} shows the reaction map for these truncated-normal diets, computed analytically at each $b$ and $P$ combination. Very small values (near zero mean movement) are colored black, showing areas of relative indifference where reaction noise will dominate.

\pagebreak
\subsection{3. Beta Distributed Percepts}
For our simulations, we bounded perceptual diets in a more natural way, by utilizing beta distributions stretched to fit $[-1,1]$. These distributions approach zero at the boundaries of our domain, fitting our asymptotic-extremes interpretation of this axis.  The beta distribution with our endpoints has the equation 
\begin{align}
    \textrm{Beta}_{[-1,1]}(x; \alpha, \beta) = 4 \frac{(1+x)^{\alpha-1} (1-x)^{\beta-1} }{2^{\alpha + \beta} } \frac{\Gamma(\alpha + \beta)}{\Gamma(\alpha) \Gamma(\beta)}\;,
\end{align}
where $\alpha$ and $\beta$ are parameters of the distribution and $\Gamma$ is the gamma function.

\begin{figure}[ht!] 
	\centering 
	\includegraphics[width=0.6\columnwidth]{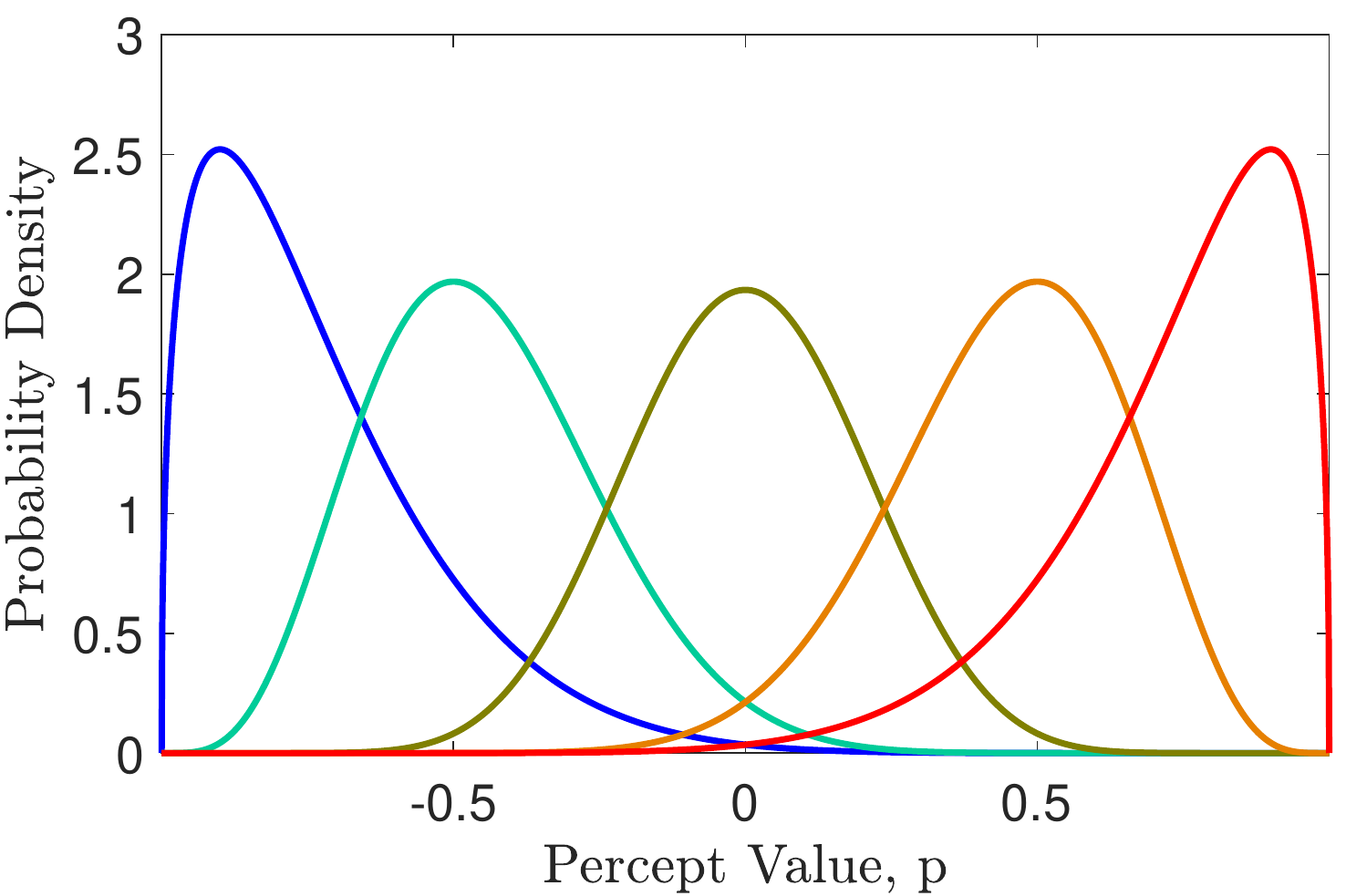}
	\caption{\textbf{Beta-distributed diets.} Examples of beta distributions with peaks at $P=-0.9$ (blue), $-0.5$ (teal), $0$ (green), $0.5$ (orange), and $0.9$ (red). All have the same standard deviation, $\sigma_p = 0.2$.}
	\label{fig:beta_diets}
\end{figure}

We can construct a distribution to have any desired mode (peak) $P=p(b)$ and standard deviation $\sigma_p$ by solving the implicit equations
\begin{align}
    \textrm{mode } = P &= \frac{\alpha - \beta}{\alpha + \beta -2} \\
    \textrm{variance } =\sigma_p^2 &= \frac{4 \alpha \beta}{(\alpha + \beta)^2(\alpha + \beta + 1)}
\end{align}
for $\alpha, \beta >1$ in terms of $P$ and $\sigma_p$. Examples are shown in Fig.~\ref{fig:beta_diets}.

\begin{figure}[ht!] 
	\centering 
	\includegraphics[width=0.63\columnwidth]{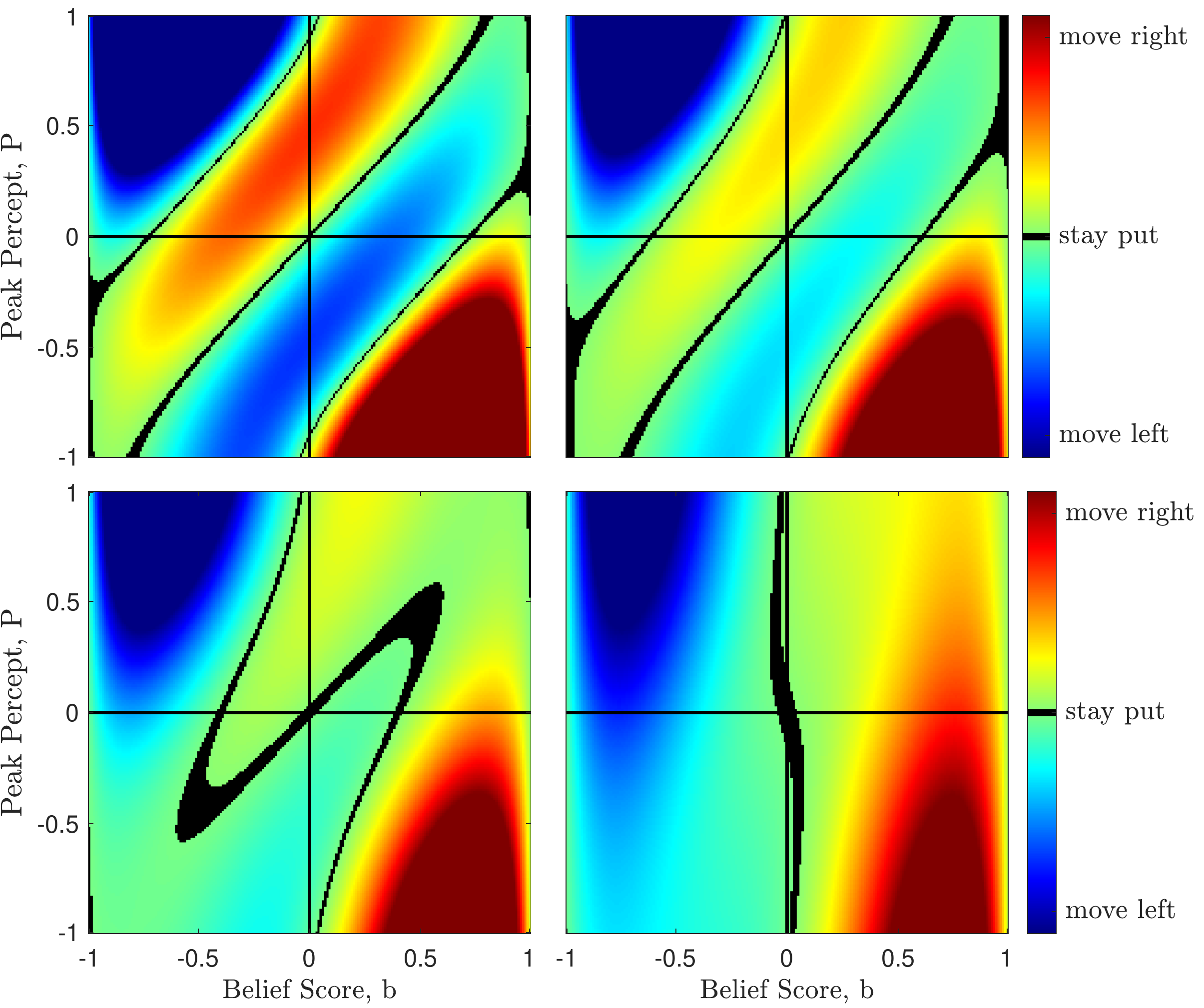}
	\caption{\textbf{Reaction map for beta-distributed diets.} Average movement caused by beta-distributed perceptual diets with peak $P$, for individuals at belief score $b$, and repulsion distance $d= 0.8$. Results shown for $\sigma_p = 0.2$ (top left),  $0.3$ (top right), $0.4$ (bottom left), and $0.5$ (bottom right).  }
	\label{fig:beta_react_map}
\end{figure}

Unfortunately, when using these beta distributions, the weighting integrals with our cubic reaction function aren't possible to evaluate in closed form. However, we may numerically compute these integrals for a finite grid of $P$ and $b$ values at any chosen standard deviation to visualize the reaction space. In Fig.~\ref{fig:beta_react_map}, we can see the repulsion boundaries bending and bifurcating as $\sigma_p$ increases.

Computing a reaction map like in Figs.~\ref{fig:unbdd_react_map}, \ref{fig:trunc_norm_react_map} or \ref{fig:beta_react_map} allows us to use our graphical analysis technique with \textit{any} perception curve, to get a sense of average population drift for the whole political spectrum.

For our ``realistic'' simulation shown in Fig.~1 of the main text, we used beta-distributed perceptual diets.  To compute in-group and out-group drift values $v_{in}$ and $v_{out}$ in a computationally feasible way, we discretized the $b$ and $P$ domains to the nearest hundredth and computed the integrals $v_{in}$ and $v_{out}$ at each possible combination---as was done for Figs.~\ref{fig:trunc_norm_react_map} and \ref{fig:beta_react_map}, which show $v_{out}$ for different $\sigma_p$ values. Then in iteration, we used nearest-neighbor interpolation with this map rather than computing each individual's weighting integral at each time-step.

\section{Appendix B: Bail et al.~In Silico Details}

For simulation of Bail's experiment \cite{Bail2018}, the population was initialized at its equilibrium, but in addition to $v_{in}$ and $v_{out}$ there was a third influence $v_{bot}$ based on an out-group distribution peaked at value $P=-0.75$ shown to Republicans and $P=0.3$ shown to Democrats, to roughly match the other party's equilibrium distribution.  This extra out-group effect was weighted as if it consisted of 24 additional percepts on top of a 100-percept daily diet, i.e., with weight $f_{bot} = 24/124$. So our full SDE becomes
\begin{align}
\tau\ \textrm{d}b &= (1-b^2) \left\{ \left[(1-f_{bot})(f v_{in} + (1-f) v_{out}) + f_{bot} v_{bot} \right] \textrm{d}t + \sigma_r \textrm{d}W \right\}\;.
\end{align}

Under this assumption, the time constant $\tau = 30$ caused movement in agreement with Bail et al.~\cite{Bail2018}: slight leftward movement of Democrat mean from $-0.51$ to $-0.53$ (about $6\%$ of its natural standard deviation), but significant rightward movement of the Republican mean from $0.30$ to $0.42$ (about $40\%$ of its natural standard deviation). 

All data and code utilized is available at https://doi.org/10.21985/n2-4msy-vj08.

\end{document}